\pdfoutput=1
\documentclass[aps,prd,amsmath,twocolumn,superscriptaddress,preprintnumbers,amssymb,showpacs,floatfix,nofootinbib,longbibliography]{revtex4-1}

\usepackage{graphicx}
\usepackage{bm}
\usepackage{hyperref}
\usepackage{slashed}
\usepackage{color}
\usepackage{aas_macros}
\usepackage{wasysym}
\usepackage{slashed}
\usepackage{lipsum}
\usepackage{subfigure}
\usepackage{multirow}
\usepackage{amsmath}
\usepackage{array}
\usepackage{varwidth}
\usepackage{lettrine}

\input Zallman.fd

\LettrineTextFont{\itshape}
\newcommand{\be}{\begin{equation}}
\newcommand{\ee}{\end{equation}}
\newcommand{\bea}{\begin{eqnarray}}
\newcommand{\eea}{\end{eqnarray}}

\hypersetup{
     colorlinks   = true,
     citecolor    = orange,
     urlcolor     = orange,
     linkcolor    = orange
}

\usepackage{listings}
\usepackage{color,xcolor}

\begin{document}

\preprint{SLAC-PUB-17594}

\title{First Analysis of Jupiter in Gamma Rays and a New Search for Dark Matter}

\author{Rebecca K. Leane}
\affiliation{SLAC National Accelerator Laboratory, Stanford University, Stanford, CA 94035, USA}

\author{Tim Linden}
\affiliation{Stockholm University and The Oskar Klein Centre for Cosmoparticle Physics,  Alba Nova, 10691 Stockholm, Sweden}

\begin{abstract}
We present the first dedicated $\gamma$-ray analysis of Jupiter, using 12 years of data from the \textit{Fermi} Telescope. We find no robust evidence of $\gamma$-ray emission, and set upper limits of $\sim10^{-9}~$GeV\,cm$^{-2}\,$s$^{-1}$ on the Jovian $\gamma$-ray flux. We point out that Jupiter is an advantageous dark matter (DM) target due to its large surface area (compared to other solar system planets), and cool core temperature (compared to the Sun). These properties allow Jupiter to both capture
and retain lighter DM, providing a complementary probe of sub-GeV DM. We therefore identify and perform a new search for DM-sourced $\gamma$-rays in Jupiter, where DM annihilates to long-lived particles, which can escape the Jovian surface and decay into $\gamma$-rays. We consequently constrain DM-proton scattering cross-sections as low as about $10^{-40}~$cm$^2$, showing Jupiter is up to ten orders of magnitude more sensitive than direct detection. This sensitivity is reached under the assumption that the mediator decay length is sufficient to escape Jupiter, and the equilibrium between DM capture and annihilation; sensitivities can be lower depending on the DM model. Our work motivates follow-up studies with upcoming MeV telescopes such as AMEGO and e-ASTROGAM.
\end{abstract}

\maketitle

\lettrine{K}{ing} of the Roman gods, Jupiter, commanded lightning, thunder, and storms. Analogous to the Greek god Zeus, he exerted his power with lightning bolts as weapons. His luminous wrath won his name one of the brightest objects in the sky, \textit{Iovis Stella} (the star of Jupiter). Today, it is known as Jupiter, which is the heaviest and largest planet in our Solar System.

For the first time, we perform a dedicated search for Jupiter's lightning bolts ($\gamma$-rays) with the \textit{Fermi} $\gamma$-ray Space Telescope. These $\gamma$-rays could potentially be produced through the active acceleration of cosmic rays in Jovian magnetic fields~\cite{2018EPJC...78..848P}, through the passive interaction of galactic cosmic rays with Jupiter's atmosphere (similar to Solar models ~\cite{Seckel:1991ffa}), or from Dark Matter (DM) annihilation. Using 12 years of \textit{Fermi} Large Area Telescope (LAT) $\gamma$-ray data, we perform a novel analysis that has been optimized for studies of solar system objects, such as the Sun~\cite{Linden:2018exo, Linden:2020lvz}. This is the first ever measurement of Jupiter in $\gamma$-rays, with important implications for our understanding of Jovian properties; see the Supplemental Material for discussion of the astrophysical implications.

Detecting, or ruling out, Jovian $\gamma$-ray emission would also have important implications for Dark Matter (DM). DM in the Galactic halo can be captured by Jupiter if it scatters with Jovian matter, loses sufficient kinetic energy, and becomes gravitationally bound. Jupiter is an advantageous DM detector for several reasons. First, compared to the Sun, it has a much cooler core. This low core temperature means that less kinetic energy is transferred to the DM, making it easier to capture and retain DM after the initial scattering. While DM evaporation inhibits Solar DM limits below a few-GeV, studies of Jupiter can probe lighter DM.  Second, compared to other planets, Jupiter is heavier and has a larger radius. This means it can capture more DM, and consequently has a larger DM annihilation rate.

\begin{figure}[t!]
    \centering
    \includegraphics[width=\columnwidth]{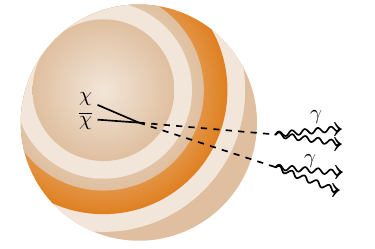}
    \caption{Schematic of DM annihilation to long-lived particles in Jupiter. The long-lived particles can decay outside the Jovian surface, producing a new source of $\gamma$-rays.}
    \label{fig:jupiter}
\end{figure}

 Figure~\ref{fig:jupiter} illustrates the DM scenario we study; to detect DM annihilation inside Jupiter, the $\gamma$-rays must escape its atmosphere. Captured DM particles annihilate to long-lived mediators that subsequently decay outside of the Jovian surface, producing a new source of $\gamma$-rays that can be detected by the \textit{Fermi} Telescope. This is the first proposed detectable signature of DM from Jupiter. We will use our new Jovian gamma-ray measurements to search for this signal for the first time.
 
Long-lived particles are theoretically well-motivated~\cite{Kobzarev:1966qya,Okun:1982xi,Holdom:1985ag,Holdom:1986eq,Batell:2009zp,Martin:1997ns}. They are currently extensively searched for in fixed-target and collider experiments~\cite{Reece:2009un,Morrissey:2014yma,Aad:2015rba}, as well as astrophysical settings~\cite{Burrows:1987zz, Burrows:1990pk, Kainulainen:1990bn, Raffelt:1996wa, Hanhart:2000er, Hanhart:2001fx, Dreiner:2003wh, Rrapaj:2015wgs, Chang:2016ntp, Chang:2018rso, Lee:2018lcj, DeRocco:2019njg, DeRocco:2019jti,Hooper:2019xss, Ertas:2020xcc,Croon:2020lrf,Bollig:2020xdr,Leane:2021ihh}. Dark sectors with long-lived mediators have previously been constrained using celestial bodies such as the Sun~\cite{Batell:2009zp,Pospelov:2007mp,Pospelov:2008jd,Rothstein:2009pm,Chen:2009ab,Schuster:2009au,Schuster:2009fc,Bell_2011,Feng:2015hja,Kouvaris:2010,Feng:2016ijc,Allahverdi:2016fvl,Leane:2017vag,Arina:2017sng,Albert:2018jwh, Albert:2018vcq,Nisa:2019mpb,Niblaeus:2019gjk,Cuoco:2019mlb,Serini:2020yhb,Mazziotta:2020foa,Bell:2021pyy}, Earth~\cite{Feng:2015hja}, and recently with large populations of brown dwarfs and neutron stars~\cite{Leane:2021ihh}. In this paper, we complement this existing parameter space, showing that Jovian $\gamma$-ray searches provide a MeV-scale cross-section sensitivity that significantly exceeds previous efforts. In light-mediator scenarios, our results are far superior to direct detection experiments.

\noindent\textbf{\textit{Fermi $\gamma$-ray Data Analysis}}--To study Jupiter in $\gamma$-rays, we perform a novel \textit{Fermi}-LAT analysis optimized for solar system objects that move with respect to the astrophysical background. The key to our method is the production of a fully \emph{data-driven} background model. We first calculate the $\gamma$-ray flux in a 45$^\circ$ region of interest (ROI) surrounding Jupiter. While this ROI vastly exceeds the $\sim$1$^\circ$ point-spread function (PSF) of \textit{Fermi}-LAT photons at $\sim$1~GeV, such a large ROI is necessary to study Jupiter at energies near 10~MeV, where the 95\% containment angle can approach 30$^\circ$. We assign every $\gamma$-ray a ``Jovian" coordinate by calculating its deviation in right ascension (RA) and declination (DEC) from the simultaneous Jupiter position. We then produce a background model by calculating the $\gamma$-ray flux at each position in equatorial coordinates during periods when Jupiter was more than 45$^\circ$ away. Finally, we determine the equatorial exposure at every pixel in the Jovian coordinate system and subtract the background flux. This produces an ``on"/``off" map that isolates Jupiter's flux and automatically accounts for astrophysical uncertainties that plague standard $\gamma$-ray analyses.

Our dataset includes all $\gamma$-rays with recorded energies between 10~MeV to 10~GeV and zenith angles below 90$^\circ$. This significantly expands on previous \textit{Fermi}-LAT studies that adopted a minimum energy of 100~MeV. Due to the large point-spread function and energy dispersion of \textit{Fermi}-LAT events between 10--100~MeV, this adds significant modeling complexity, which we address below.  

For each recorded $\gamma$-ray, we calculate its offset (in RA, DEC) from the simultaneous positions of Jupiter, the Sun and the Moon. We bin the data into 60 energy bins (20 logarithmically-spaced bins per decade), and calculate the exposure across the entire sky in 1~hr (3600-second) increments. Over this period, Jupiter moves only 0.003$^\circ$ with respect to the equatorial coordinate system, while the Sun moves $\sim$0.04~$^\circ$ and the Moon moves 0.5$^\circ$. These shifts are small compared to our ROIs and the instrumental PSF, justifying our treatment of each source as stationary \emph{within} each time bin.

\begin{figure*}[t!]
    \centering
    \includegraphics[width=2\columnwidth]{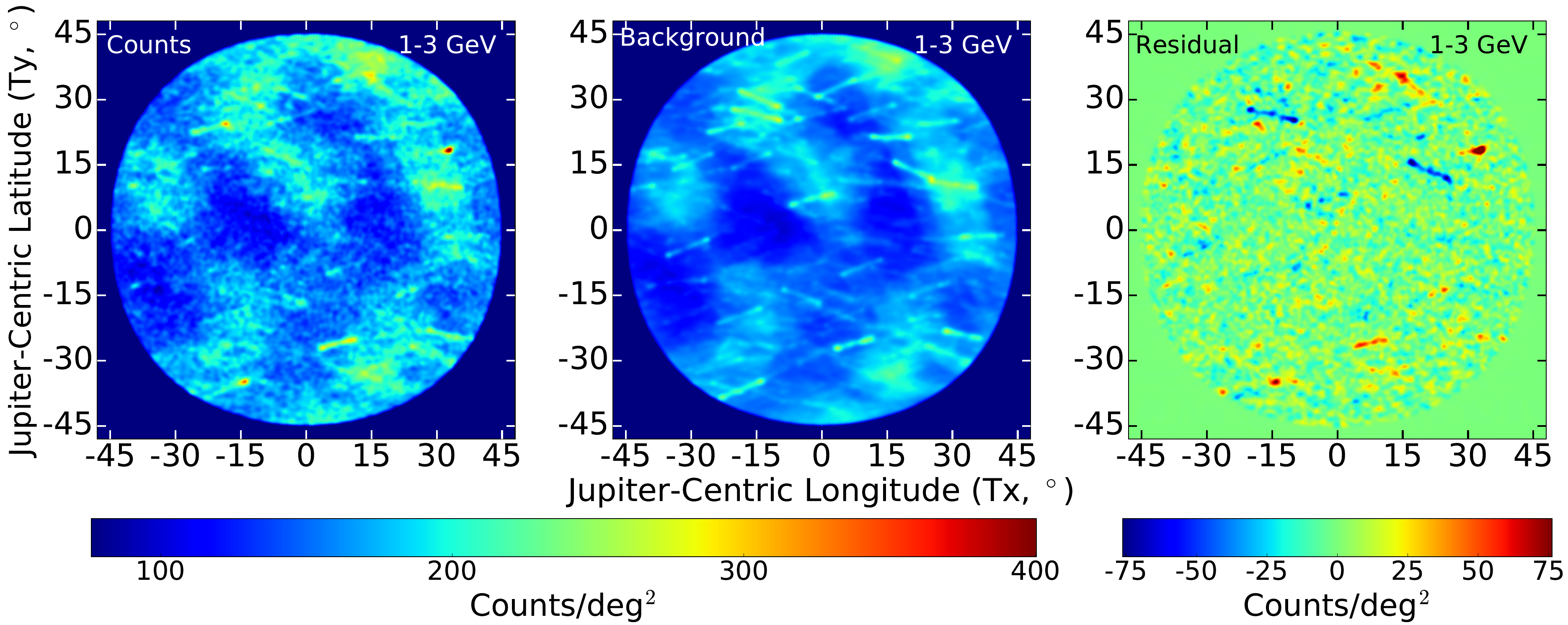}
    \caption{\textit{Fermi}-LAT $\gamma$-ray data utilized in our analysis. For the visualization of this figure, we combine all energy bins between 1$-$3.16~GeV, and smear all results with a 1$^\circ$ Gaussian, choices that are not made in the analysis of our data.
    \textbf{Left:} Counts map produced by all events recorded within 45$^\circ$ of the position of Jupiter. \textbf{Middle:} The background map, which is produced from observations of events calculated by examining identical regions in RA/DEC when Jupiter is not present. \textbf{Right:} The residual calculated by subtracting the model from our data. }
    \label{fig:jupdata}
\end{figure*}

\begin{figure}[t!]
    \centering
    \includegraphics[width=\columnwidth]{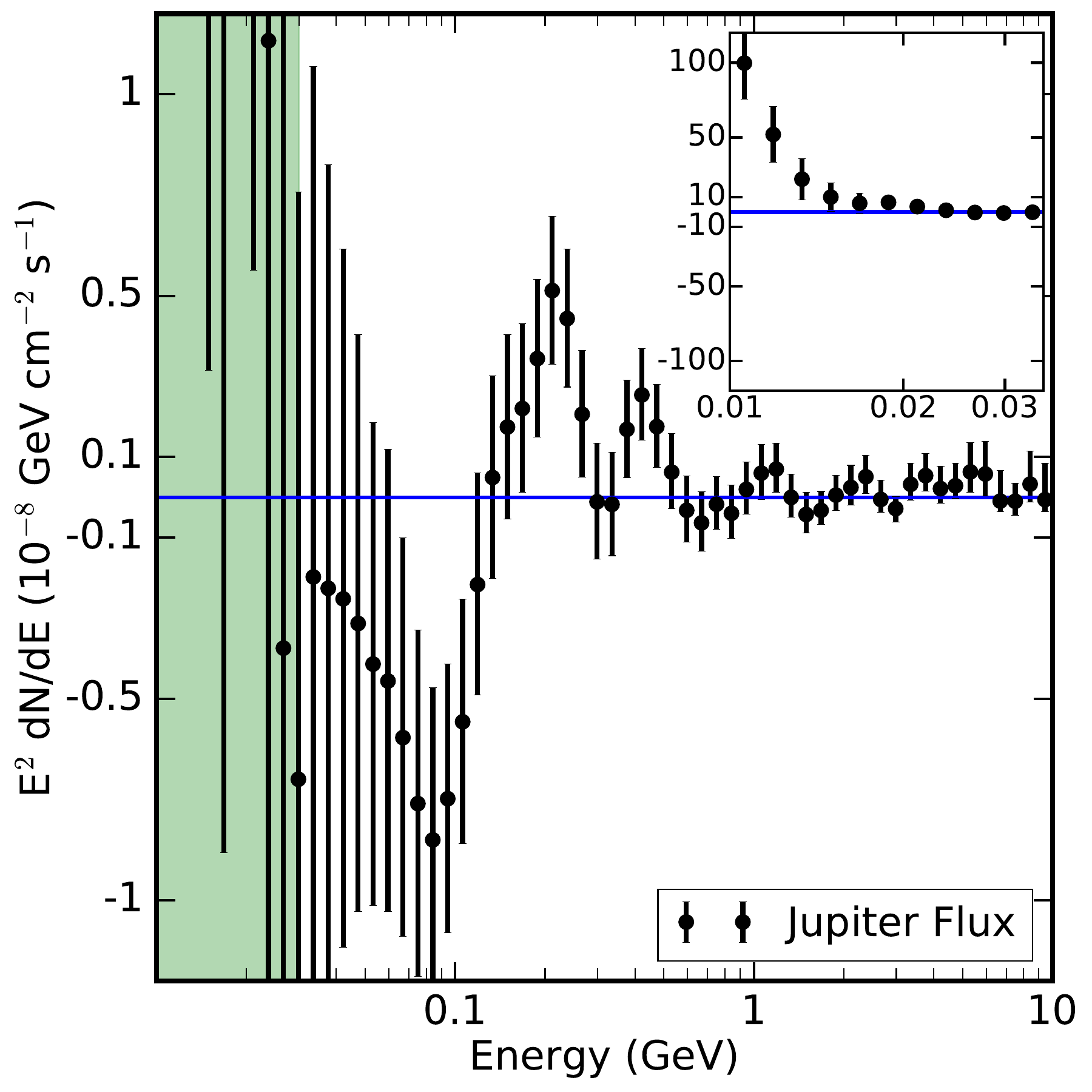}
    \caption{The $\gamma$-ray flux from Jupiter obtained in our analysis. The blue horizontal line depicts no $\gamma$-ray flux. The significant energy dispersion (especially at low-energies) makes the flux in nearby energy bins highly correlated. Through most of the energy range, we find no evidence for Jovian $\gamma$-ray emission.  In the inset (green region), we zoom to show the bright emission in the lowest energy bins.}
    \label{fig:jupiterflux}
\end{figure}

To build our background model, we remove all photons recorded within 45$^\circ$ of Jupiter (our ``on" region), within 40$^\circ$ of the Sun (which has an extended halo~\cite{2011ApJ...734..116A}), and within 20$^\circ$ of the Moon (which produces only disk emission). While the solar and lunar flux could be modeled and fit in the analysis (this approach was taken for lunar emission in Ref.~\cite{Linden:2020lvz}), this adds significant complexity because the Sun and Moon are much brighter than Jupiter. In this analysis, we simply mask these sources, losing only $\sim$15\% of the total Jupiter exposure. We also remove bright flares that approach too close to Jupiter, see the Supplemental Material for details.

Using this background model we calculate the $\gamma$-ray photon count at each RA/DEC during periods when Jupiter is not present. This is possible because Jupiter is far from any single RA/DEC most of the time. Because we have also calculated the exposure in equatorial coordinates in fine temporal bins, we can translate the photon count into a background $\gamma$-ray flux. Finally, in every 3600-second window, we convert each point in equatorial coordinates to its simultaneous position in Jovian coordinates, producing an entirely data-driven background model at each point in Jovian coordinates. Finally, we model Jupiter itself. Because the spatial extent of Jupiter is much smaller than the \textit{Fermi} PSF at any energy, we treat Jupiter as a point-source, details are in the Supplemental Material.

Figure~\ref{fig:jupdata} shows our model at energies between 1--3~GeV, including the $\gamma$-ray flux within 45$^\circ$ of Jupiter, the $\gamma$-ray flux predicted by our background model, and the resulting residual. Bright lines across the ROI correspond to bright sources moving through the Jovian coordinate system. The residuals are generally only a few percent, with maximum values near 20\%. These primarily stem from flaring sources that we did not remove. The scale of Jupiter in this 1--3~GeV energy bin is about 1/45th of the image width. The contribution from Jupiter will cover a much larger region in the lowest-energy bins, justifying our usage of a full 45$^\circ$ ROI.\\

\noindent\textbf{\textit{Upper Limits on the Jovian $\gamma$-ray Flux}}--We utilize \texttt{iminuit} to calculate the simultaneous fit of our background model and the Jupiter flux in each energy bin. In this case, we utilize a simple two-parameter fit, where the normalization of the Jupiter flux, and the normalization of the overall background template, are allowed to float independently in each energy bin. We note two important details. First, the normalization of the background template should equal 1, as the normalization of background sources should be independent of the position of Jupiter. Indeed, we find only very small deviations (on the order of $1-2$\%) from unity, verifying the accuracy of our techniques. Small errors may stem from variable sources, or instrumental exposure corrections that are correlated with the Jupiter position. 

We allow the normalization of Jupiter to assume both positive or negative values. This is important, because constraining the Jupiter flux to be positive (and binning the data finely in energy) may make upward fluctuations appear overly significant. However, this choice can add complexity in the high-energy regime, because Poisson statistics are ill-defined when the total model expectation is negative in any pixel. Here, we follow Ref.~\cite{Linden:2019soa}, calculating the Poisson statistic from its absolute value in bins where the observed number of counts is 0, but ruling out negative fluctuations in bins where the number of observed photons is non-zero. This choice is numerically important, but has no practical impact on our results.

Figure~\ref{fig:jupiterflux} shows the Jovian $\gamma$-ray flux in our analysis. We note several important results. First, the overall flux of Jupiter is consistent with 0. For an E$^{-2}$ $\gamma$-ray spectrum we obtain a 95\% confidence upper limit on the Jovian energy flux of 9.4$\times$10$^{-10}$~GeV~cm$^{-2}\,$s$^{-1}$ between \mbox{10~MeV -- 10~GeV}, while for a cosmic-ray motivated spectrum of E$^{-2.7}$ we obtain an upper limit of 3.2$\times$10$^{-9}$~cm$^{-2}\,$~s$^{-1}$. For power-law spectra between E$^{-1.5}$ and E$^{-3.0}$, the significance of Jovian emission never exceeds 1.5$\sigma$. Second, we note that the error bars in our analysis are highly correlated. This is due to the significant energy-dispersion of low-energy \textit{Fermi}-LAT data, which smears the true Jovian energy flux between multiple energy bins. This effect decreases, from $\sim$30--50\% in the lowest energy bins, to near 15\% at GeV energies. Third, we note that there is a statistically significant excess in the lowest energy bins (below 15~MeV). The local significance of this excess is 4.6$\sigma$ in the energy bin between 10---11.2~MeV, 2.3$\sigma$ in the bin between 11.2---12.6~MeV, and 1.3$\sigma$ in the bin between 12.6---14.1~MeV. Combined, these provide a $5\sigma$ local excess. This is a potentially exciting result, pointing to the possibility that Jupiter may be capable of accelerating cosmic-rays to MeV-energies in its strong electromagnetic fields. However, significant caution is warranted. Firstly, this analysis severely pushes the limits of the \textit{Fermi}-LAT. To our knowledge, no other study of steady-state emission has taken place in such a low-energy regime. Numerous systematic effects may be present in the low-energy bins that would be difficult to control in any analysis, and a detailed study of systematics in this region (which lies beyond the scope of this paper), would be necessary. Secondly, the \textit{Fermi}-LAT effective area rises rapidly with energy in this regime. The exposure at 20~MeV is 50 times larger than at 10~MeV. The fact that no excess is observed in the 20~MeV energy bins strongly constrains the spectrum of any 10~MeV excess. Effectively, any power-law emission at 10~MeV (with spectra harder than $\sim$E$^{-3}$) is ruled out, and the emission observed at 10~MeV must have a spectrum that is strongly exponentially suppressed.
Fourth, we find a low-significance excess (2$\sigma$ local), best fit by the annihilation of a DM particle of mass 493~MeV into long-lived particles which decay directly into $\gamma$-rays. However we do not consider this sufficiently statistically significant.\\

\noindent\textbf{\textit{Dark Matter in Jupiter}}--DM from the Galactic halo can fall into Jupiter, scattering and losing energy. Once the kinetic energy of the DM is less than the gravitational potential, the DM particle is captured. DM capture can occur via single or multiple scatters with Jovian matter \cite{Kouvaris:2010,Bramante:2017,Dasgupta:2019,Ilie:2020}. The DM capture rate for $N$ required scatters is given by~\cite{Bramante:2017}
\begin{align}
C_N &= \pi R_{\jupiter}^2 p_N(\tau) \frac{\sqrt{6} n_\chi}{3 \sqrt{\pi} \bar{v}} \times\\ &\left((2 \bar{v}^2 + 3 v_{\rm esc}^2) - (2 \bar{v}^2 + 3 v_N^2)\exp \left(-\frac{3(v_N^2 - v_{\rm esc}^2)}{2 \bar{v}^2}\right)   \right),\nonumber
\label{eq:multi scatter}
\end{align}
where $v_{\rm esc} = \sqrt{{2\,G\,M_{\jupiter}}/{R_{\jupiter}}}$ $\sim60\,$km/s is Jupiter's escape velocity with $G$ as the gravitational constant, $M_{\jupiter}=1.9\times 10^{27}$~kg and $R_{\jupiter}=69,911$~km are the mass and radius of Jupiter respectively. $\bar{v}$ is the DM velocity dispersion, $n_\chi(r)$ is the DM number density at the Jupiter position, related to the mass density via $n_\chi(r)=\rho(r)/m_\chi$, and $v_N = v_{\rm esc} (1 - \beta_{+}/2)^{-N/2}$ with $\beta_{+} = {4 m_{\chi} m_n}/{(m_{\chi} + m_n)^2}$. Note that here we have assumed that a scattering variable $z=\textrm{sin}^2(\theta_{\rm CM}/2)$, where $\theta_{\rm CM}$ is the CoM scattering angle, takes its average value of $\langle z_i\rangle=1/2$, which is not a perfect assumption for the single scatter limit, but is accurate within a factor of a few in our case. The probability of a single DM particle undergoing $N$ scatters is
\begin{equation}
    p_N(\tau) = 2 \int_0^1 dy \frac{y e^{-y \tau} (y \tau)^N}{N!},
\end{equation}
where $y$ is the cosine of the incidence angle
of DM entering Jupiter, and $\tau$ is the optical depth,
\begin{equation}
    \tau=\frac{3}{2}\frac{\sigma}{\sigma_{\rm sat}},
\end{equation}
and $\sigma_{\rm sat}$ is the saturation cross section of DM capture onto nucleons given by $\sigma_{\rm sat} = {\pi R^2}/{N_n}$, where $N_n$ is the number of Jovian nucleons. We assume for simplicity that Jupiter is $100\%$ hydrogen. The total capture rate of DM in Jupiter $C_{\jupiter}$ is then given by
\begin{equation}
    C_{\jupiter} = \sum_{N= 1}^{\infty} C_N.
\label{eq:multiscatter_total}
\end{equation}
Assuming that equilibrium between capture and annihilation of DM within Jupiter is reached, the annihilation rate ($\Gamma_{\rm ann}$) is simply $\Gamma_{\rm ann} =C_{\jupiter}/2$.

We assume that DM annihilates into two mediators $\phi$ that have a sufficiently long lifetime $\tau$ or large boost factor $\gamma \approx m_\chi/m_\phi$ such that the decay length $L$ exceeds the radius of Jupiter $R_{\jupiter}$, as
\begin{equation}
L = \gamma \beta \tau \simeq \gamma c \tau > {R_{\jupiter}}.
\label{eq:decay}
\end{equation}
The total flux at Earth from long-lived particles in Jupiter is given by~\cite{Leane:2017vag}
\begin{equation}
E^2\frac{d\Phi}{dE} = \frac{\Gamma_{\rm ann}}{4\pi D_\oplus^2}\times E_\gamma^2\frac{dN_\gamma}{dE_\gamma} \times \mathrm{BR(X\rightarrow SM)} \times P_{\rm surv},
\label{eq:flux}
\end{equation}
where $D_\oplus$ is the average distance of Jupiter to Earth, $\mathrm{BR(X\rightarrow SM)}$ is the branching ratio of the mediator to a given SM final state. The probability of the signal surviving to reach the detector near Earth, $P_{\rm surv}$, provided the decay products escape Jupiter is~\cite{Leane:2017vag}
\begin{equation}
    P_{\rm surv} = e^{-{R_{\jupiter}}/\gamma c \tau} - e^{-D_\oplus/ \gamma c \tau}.
\end{equation}
In Eq.~\ref{eq:flux}, the $E_\gamma^2\,dN_\gamma/dE_\gamma$ term corresponds to the $\gamma$-ray energy spectrum. The relevant DM annihilation process is $\chi\overline{\chi}\rightarrow\phi\phi\rightarrow4\,{\rm SM}$. DM annihilation to two mediators is dominant over DM annihilation to one mediator, as it is not phase-space suppressed. This yields the characteristic $\gamma$-ray box spectral shape~\cite{Ibarra:2012dw}. As we consider mediators at least a factor few lighter than the DM, the highest energy $\gamma$-rays always peak close to the DM mass. This means that our results are approximately independent of the mediator mass (provided it is sufficiently boosted/long-lived to escape Jupiter).

\begin{figure}[t!]
    \centering
    \includegraphics[width=\columnwidth]{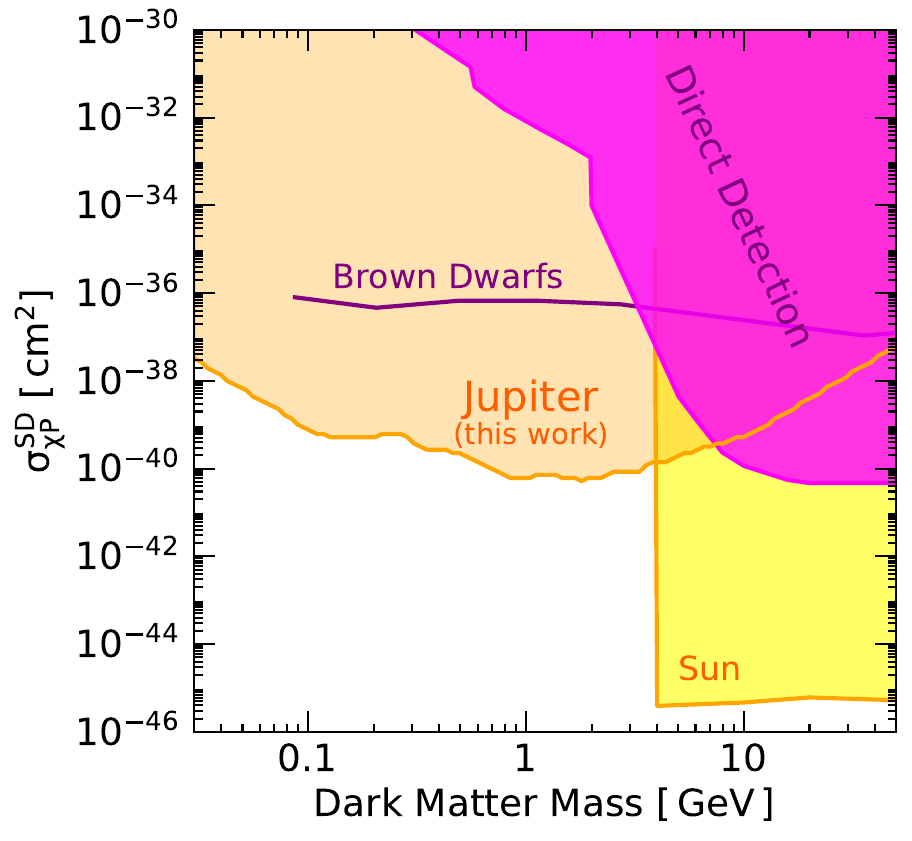}
    \caption{95$\%$ C.L. DM-proton scattering cross section limits as a function of DM mass $m_\chi$, arising from DM annihilation to long-lived particles, from our new Jovian $\gamma$-ray search. We show complementary constraints from direct detection~\cite{Gangi:2019zib, Aprile:2019dbj, Aprile:2019jmx,Liu:2019kzq}, as well as DM annihilation to long-lived particles in the Sun~\cite{Leane:2017vag,Albert:2018jwh}, and brown dwarfs in the Galactic center~\cite{Leane:2021ihh}.}
    \label{fig:limits}
\end{figure}

Figure~\ref{fig:limits} shows our new  cross section constraints on DM annihilation to long-lived particles using Jovian $\gamma$-rays at 95$\%\,$C.L., for mediator decay to $\gamma$-rays, via $\chi\chi\rightarrow\phi\phi$, $\phi\rightarrow2\,\gamma$. In this plot we take the mediator to decay at the Jovian surface. We show for comparison, limits from direct detection (DD)~\cite{Gangi:2019zib, Aprile:2019dbj, Aprile:2019jmx,Liu:2019kzq}, which loses sensitivity with lower DM masses as the recoils become increasingly weak. Jupiter on the other hand, is optimized to search for DM particles with masses of around the proton mass, providing up to 10 orders of magnitude stronger sensitivity than DD. While we only show limits for direct decay to gamma rays, our search is also sensitive to other final states, which produce gamma rays via electromagnetic bremsstrahlung or hadronic decays. As we show constraining power of up to 10 orders of magnitude higher than previous limits, we expect that while other final states can produce weaker limits, there should still be significant constraining gain. Note that compared to the Jupiter limits, the DD limits do not require any minimum annihilation cross section. We also show complementary searches for DM and long-lived particles in the Sun~\cite{Leane:2017vag,Albert:2018jwh}, and the Galactic Center (GC) population of brown dwarfs (BDs)~\cite{Leane:2021ihh}. Compared to GC BDs, Jupiter is only one object in a comparably low DM density, however it is very close to Earth, leading to superior sensitivities. Compared to solar DM results as calculated in Refs.~\cite{Leane:2017vag,Albert:2018jwh}, the solar limits extend to much lower cross sections, owing to the Sun being much larger than Jupiter and closer to the Earth. The Jupiter limits extend to lower DM masses, because Jupiter’s cooler core in part can more easily prevent evaporation of sub-GeV DM. However, depending on the specific DM model of interest, we emphasize that the low-mass end of the sensitivity can substantially change.  In the case of only heavy mediators, the DM evaporation mass is about 200 MeV - 1 GeV for Jupiter depending on the scattering cross section. In the case of light mediators, the DM evaporation mass can be instead at least sub-MeV~\cite{Acevedo:2023owd}. We therefore show the sensitivity of Fermi assuming no evaporation, but the exact lower bound where the sensitivity truncates will depend on the DM model. Discussion of some example models is given in Sec.~\ref{sec:models} of the Supplemental Material.\\

\noindent\textbf{\textit{Conclusions}}--We produced the first ever measurement of Jupiter in $\gamma$-rays, using 12 years of data from the \textit{Fermi} $\gamma$-ray Space Telescope. Our results are important for understanding Jupiter's atmospheric properties and magnetic fields, and furthermore apply to a new DM signature. We designed a new analysis framework that led to the first \textit{Fermi} steady-state analysis down to 10 MeV. This was made possible as all instrumental uncertainties (point-spread function, energy dispersion, effective area) are directly accounted for in our data-driven background model. This unlocks the power of our low-energy data, where these uncertainties become particularly acute. 

Across most $\gamma$-ray energies, we found no $\gamma$-ray flux in excess of background expectations, setting the first upper limits on the Jovian $\gamma$-ray flux. At lower $\gamma$-ray energies, we find statistically significant evidence for Jovian $\gamma$-ray emission below 15~MeV at $5\sigma$ local. While this emission has an extremely soft spectrum and is not well fit by any DM model, it may provide significant evidence of primary cosmic-ray acceleration within the Jovian atmosphere. However, this analysis pushes the envelope of \textit{Fermi}-LAT's sensitivity as an MeV $\gamma$-ray detector, and should not yet be taken as robust. We emphasize the need for new, and robust analyses for MeV Jovian $\gamma$-rays, which can be provided by proposed MeV telescopes such as AMEGO or e-ASTROGAM.

We pointed out that Jupiter is an ideal DM detector. Compared to the nearby Sun, Jupiter has a cooler core, which can prevent the evaporation of lighter DM particles, allowing new sensitivity to sub-GeV DM. Compared to other solar system planets, Jupiter is much larger, allowing a larger capture and consequent annihilation rate. We showed that if captured DM annihilates to sufficiently long-lived/boosted mediators, the mediators can escape the Jovian surface, and decay into $\gamma$-rays that are detectable by the \textit{Fermi} Telescope. We find a low-significance excess, best fit by the annihilation of a DM particle of mass 493~MeV into long-lived particles which decay directly into $\gamma$-rays. However, the local significance of this excess only slightly exceeds 2$\sigma$. We therefore used our new upper limits on the Jovian flux to constrain, for the first time, the annihilation of DM to long-lived mediators in Jupiter, for DM with masses above a few tens of MeV, with DM-proton scattering cross sections down to about $10^{-40}\,$cm$^{-1}$. This is up to 10 orders of magnitude more sensitive than DD. We emphasize, however, that the lower end of the DM mass sensitivity and cross section limits can weaken, particularly in the context of specific particle-models. These limits should instead be interpreted as demonstrating the strong constraining power of this search, rather than generic, robust constraints. Our results motivate model-dependent studies of the DM parameter space that can be constrained using Jovian $\gamma$-rays.
\\

\noindent\textbf{\textit{Acknowledgments--}}We thank John Beacom, Joe Bramante, Regina Caputo, Colleen Kron, Nestor Mirabal, Payel Mukhopadhyay, Annika Peter, Juri Smirnov, Natalia Toro, and Bei Zhou for helpful discussions and comments. RKL is supported in part by the U.S. Department of Energy under Contract DE-AC02-76SF00515. TL is partially supported by the Swedish Research Council under contract 2019-05135, the Swedish National Space Agency under contract 117/19 and the European Research Council under grant 742104. This project used computing resources from the Swedish National Infrastructure for Computing (SNIC) under the project No.~2020/5-463 partially funded by the Swedish Research Council through grant agreement no. 2018-05973.

\clearpage
\newpage
\maketitle
\onecolumngrid
\begin{center}
\textbf{\large First Analysis of Jupiter in Gamma Rays and a New Search for Dark Matter}

\vspace{0.05in}
{ \it \large Supplementary Material}\\ 
\vspace{0.05in}
{Rebecca K. Leane and Tim Linden}
\end{center}
\onecolumngrid
\setcounter{equation}{0}
\setcounter{figure}{0}
\setcounter{section}{0}
\setcounter{table}{0}
\setcounter{page}{1}
\makeatletter
\renewcommand{\theequation}{S\arabic{equation}}
\renewcommand{\thefigure}{S\arabic{figure}}
\renewcommand{\thetable}{S\arabic{table}}

The Supplementary Material contains additional details relevant to our searches, supporting the conclusions of the main text.

\tableofcontents

\newpage

\section{Additional Results for the Jovian Gamma-Ray Analysis}
\label{sec:appjov}

In this section, we show our Jupiter maps for lower energy $\gamma$-rays than the main text, in the region where the statistically significant excess occurs. We provide additional analysis details, including the ROI cuts to remove bright flares or sources. We also repeat the analysis of the main text with larger energy bins, finding consistent results.

\subsection{Jupiter Maps in MeV Gamma-Rays}

\begin{figure*}[b!]
    \centering
    \includegraphics[width=0.85\columnwidth]{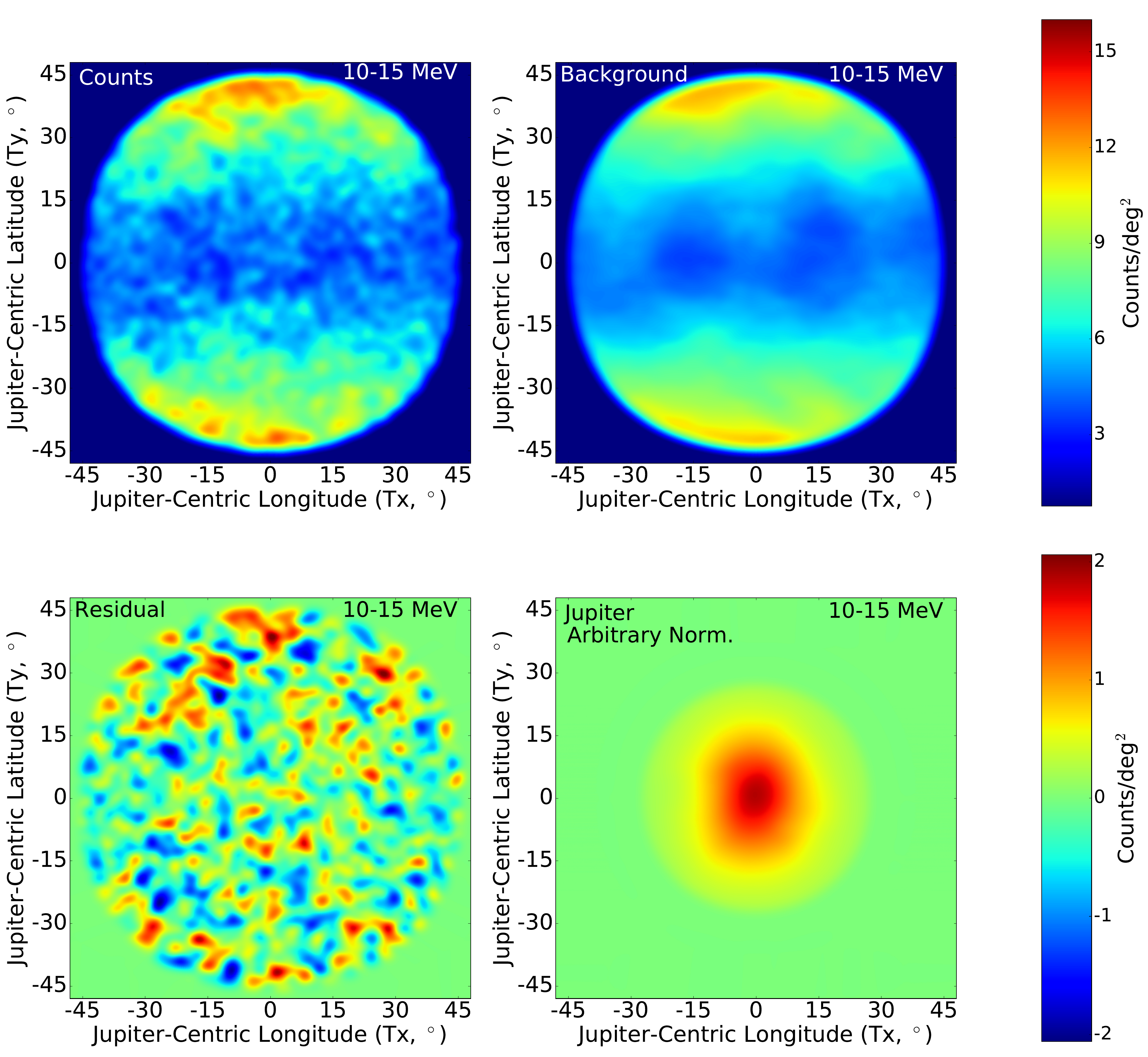}
    \caption{Same as Figure~2 of the main text, except for the energy range between 10--15~MeV, where the \textit{Fermi} effective area is small and the angular resolution of the instrument is relatively poor (and also includes the size of Jupiter at these energies for scale in the bottom-right panel). Results are instead smeared with a 3$^\circ$ Gaussian due to the low photon counts. No obvious errors in the background model appear even at these very low energies.}
    \label{fig:jupdata_lowenergy}
\end{figure*}

Figure~\ref{fig:jupdata_lowenergy} shows the same data panels as in Fig.~\ref{fig:jupdata}, but in the much lower energy range of 10--15~MeV. This lies at the extreme edge of the \textit{Fermi}-LAT energy range, creating two significant effects. First, the effective area is very small, producing a small photon count and large statistical fluctuations. Second, the PSF is poor, with an 68\% containment region of $\sim$15$^\circ$. Despite these significant issues, our background model fits the data reasonably well. Notably the residual map appears consistent with Poisson fluctuations. Interestingly, we find several upward fluctuations in the vicinity of Jupiter itself, and show in the main text that these residuals translate into a relatively significant excess consistent with Jovian emission. 

Compared to Fig.~\ref{fig:jupdata}, Fig.~\ref{fig:jupdata_lowenergy} also includes the size of Jupiter at these energies for scale in the bottom-right panel. Note that compared to scale of Jupiter in the energy range of Fig.~\ref{fig:jupdata}, the scale of Jupiter here is much larger. The aspherical appearance of Jupiter in the bottom-right panel is due to ROI cuts from masked sources, which are asymmetric in the Jovian coordinate system.

\subsection{Additional Analysis Details}

For our dataset, we place standard instrumental cuts using {\tt gtmktime}.
For the exposure calculation, we utilize {\tt phibins=10}, which is the number of bins for the azimuthal co-ordinate $\phi$. This is a non-standard choice which corrects for the fact that Jupiter's orbit through our solar system gives it a non-standard distribution in the instrumental coordinate $\phi$, because the $\phi$ distribution is strongly biased by the solar position. 

Because the spatial extent of Jupiter is much smaller than the \textit{Fermi} PSF at any energy, we treat Jupiter as a point-source, and utilize the tool \emph{gtpsf} to calculate the effective PSF of Jupiter in each of our 60 energy bins and 3600-second time bins. We calculate the average PSF by performing a weighted average of each PSF over the simultaneous Jupiter exposure.  While our background model automatically includes the effect of energy dispersion, we model the dispersion of Jupiter using the updated 15-parameter fit for P8R3 data described in Ref.~\cite{website}.

\subsection{Treatment of Flares and Details of ROI Cuts}

Jupiter is a very dim $\gamma$-ray source, which has an orbital period of 12 years, and thus has only moved through the equatorial coordinate system once. Therefore it is necessary to perform spatial and temporal cuts on any bright flares or sources that approach too close to Jupiter. Indeed, as the background model and source region are produced from photons recorded at different times, our model is sensitive to time-variable sources. This is different to solar analyses~\cite{Linden:2020lvz}, where this effect is significantly lessened for two reasons: (1) the Sun is bright, and its total luminosity in helioprojective coordinates dominates even the brightest flares, (2) the Sun moves through the equatorial coordinate system annually, and thus any variable source has been sampled (and averaged over) a dozen times. To account for this effect, we also remove regions surrounding bright, variable background sources that lie near Jupiter.

Table~\ref{tab:removedsources} shows our ROI cuts chosen to minimize backgrounds. Utilizing the \textit{Fermi} All-Sky Variability Analysis Catalog~\cite{2017ApJ...846...34A}, we mask an ROI surrounding any source that ever approaches within 45$^\circ$ of Jupiter, and which has a peak-weekly flux (during any week, regardless of its proximity to Jupiter) exceeding 3$\times$10$^{-6}$ ph~cm$^{-2}\,$s$^{-1}$. We remove a 5$^\circ$ ROI unless the source approaches to within 10$^\circ$ of Jupiter, in which case we remove a 10$^\circ$ ROI. We also remove a region around Geminga, which is not variable, but is extremely bright.

\begin{table}[h!]
  \centering
\begin{tabular}{| c | c | c | c |}
\hline
 Source & R.A. & Dec. & ROI ($^\circ$) \\ \hline \hline
Sun & --- & --- & 40 \\
Moon & --- & --- & 20 \\
Crab & 83.63 & 22.01 & 10 \\
3C 279 & 194.05 & -5.79 & 10 \\
3C 273 & ~187.28~ & 2.05 & 10 \\
~PKS 1830-211~ & 278.42 & ~-21.06~ & 10 \\
3C 454.3 & 343.49 & 16.15 & 5 \\
~PKS 2247-131~ & 342.50 & -12.86 & 5 \\
~PKS 1329-049~ & 203.02 & -5.16 & 5 \\
Geminga & 98.48 & 17.77 & 5 \\
\hline
\end{tabular}
  \caption{Regions around bright sources that are removed from our analysis. These include sources that move with respect to the equatorial coordinate system (Sun, Moon), bright variable sources that enter the Jupiter ROI, and Geminga, which is not variable, but is very bright. Removing these ROIs decreases the exposure of Jupiter by $\sim$20\%.}
  \label{tab:removedsources}
\end{table}

In the final analysis of Jupiter, we choose a large region of interest in order to make sure the background model is well-constrained by the emission in regions very far from the position of Jupiter. That is, the analysis ROI must be much bigger then the size of the source we are analyzing. Our choice of a 45$^\circ$ ROI is large, and very conservative -- motivated by the lowest energy bins of our analysis, where the PSF of the instrument can approach $10-15^\circ$.

The ROI to exclude variable sources (or the Sun or Moon), on the other hand -- does not need to exceed the size of the source itself (because we are not attempting an analysis in this region, but are only attempting to get rid of any unexpected contributions from the source). Thus the masked region does not need to exceed the source size. Additionally, we must make a choice to balance the efficient removal of stray photons from each source, with a choice to maintain as much exposure on Jupiter as possible. To this effect, we note that a mask that removes 80\% of a sources photons decreases the effect of a variable source on our model by 80\% (some photons leak out of the mask both into both the ``on" and ``off" regions that are used to determine the Jupiter flux and background model). Thus, for this analysis, we lean towards making a slightly stricter cut (removing $\sim$100\% of the variable source photons above $\sim$300~MeV, but potentially only removing $\sim70-90\%$ of the photons at lower energies), while still maintaining a significant exposure for Jupiter.\\

\subsection{Analyses with Coarse Energy Bins}

A unique advantage of our Jovian analysis is the ability to utilize the relatively few astrophysical background and Jovian templates, in order to analyze the Jovian $\gamma$-ray flux in very fine energy bins, without losing power to statistical fluctuations or model degeneracies stemming from low-photon counts. The choice of final spatial binning is well motivated for our dark matter searches, which produce very hard $\gamma$-ray spectra with significant energy deposition in the bin directly below the dark matter mass.

In order to take advantage of this energy resolution, the effect of energy-dispersion must be carefully accounted for, as the \textit{Fermi}-LAT instrument has an imperfect energy resolution that smears the true energy of recorded events. This dispersion is only $\sim$5\% at energies near 10~GeV, but is nearly 30\% for the 10 MeV events in our low-energy analysis. Thus, at low energies, the energy dispersion is much larger than the energy binning -- and can move events between energy bins. Throughout the paper, we have carefully tracked this effect, by taking the true energy of events and dispersing them utilizing the modeled \texttt{P8\_V3} \textit{Fermi}-LAT instrument response functions.

In this Appendix, we instead repeat the analysis, but utilize larger energy bins to minimize the effects of dispersion. In particular, we produce models employing 10 and 5 logarithmic energy bins per decade (instead of 20 energy bins, as in the main text). All other parameters (including energy dispersion between bins) are treated identically. 

\begin{figure*}[b!]
    \centering
    \includegraphics[width=0.45\columnwidth]{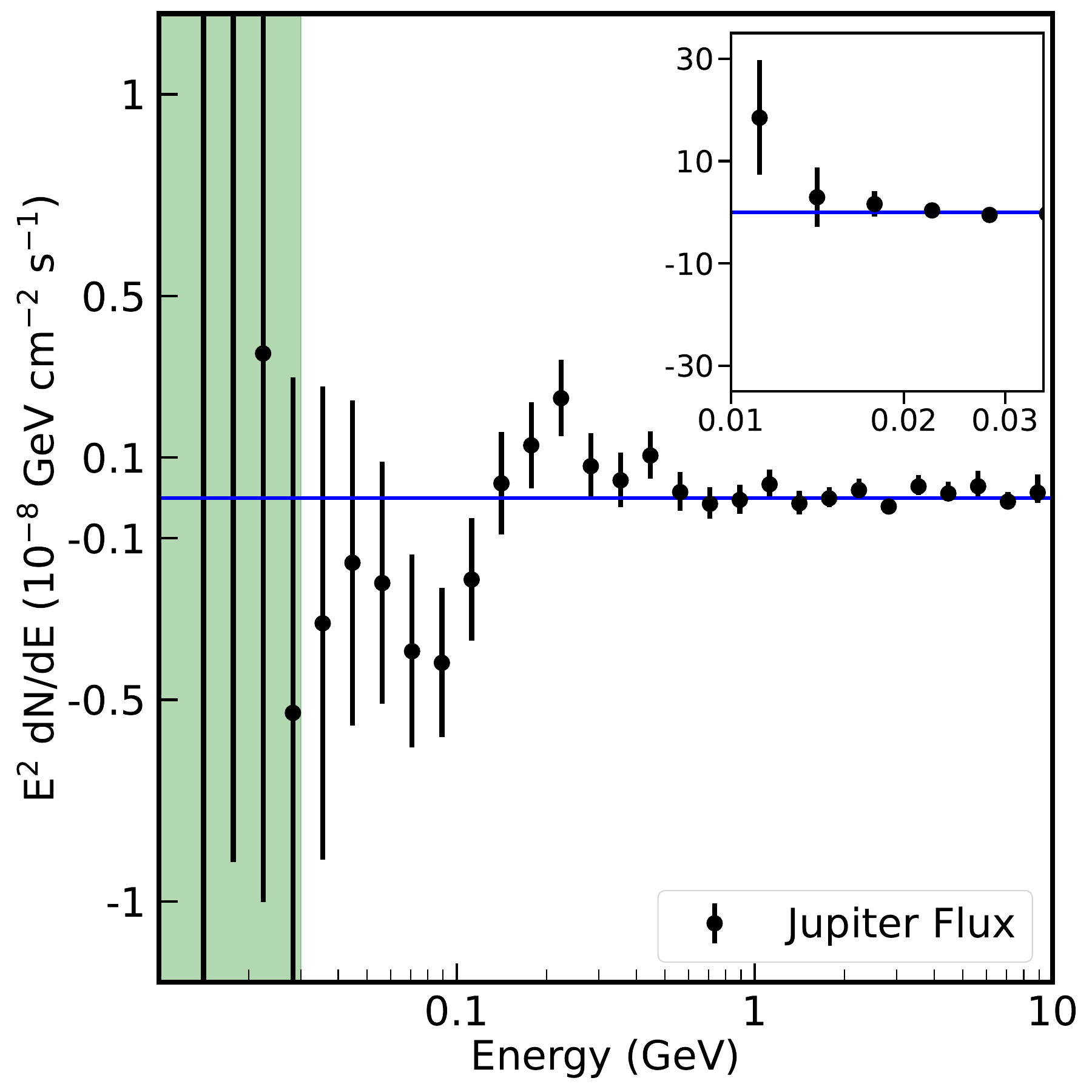}
    \includegraphics[width=0.45\columnwidth]{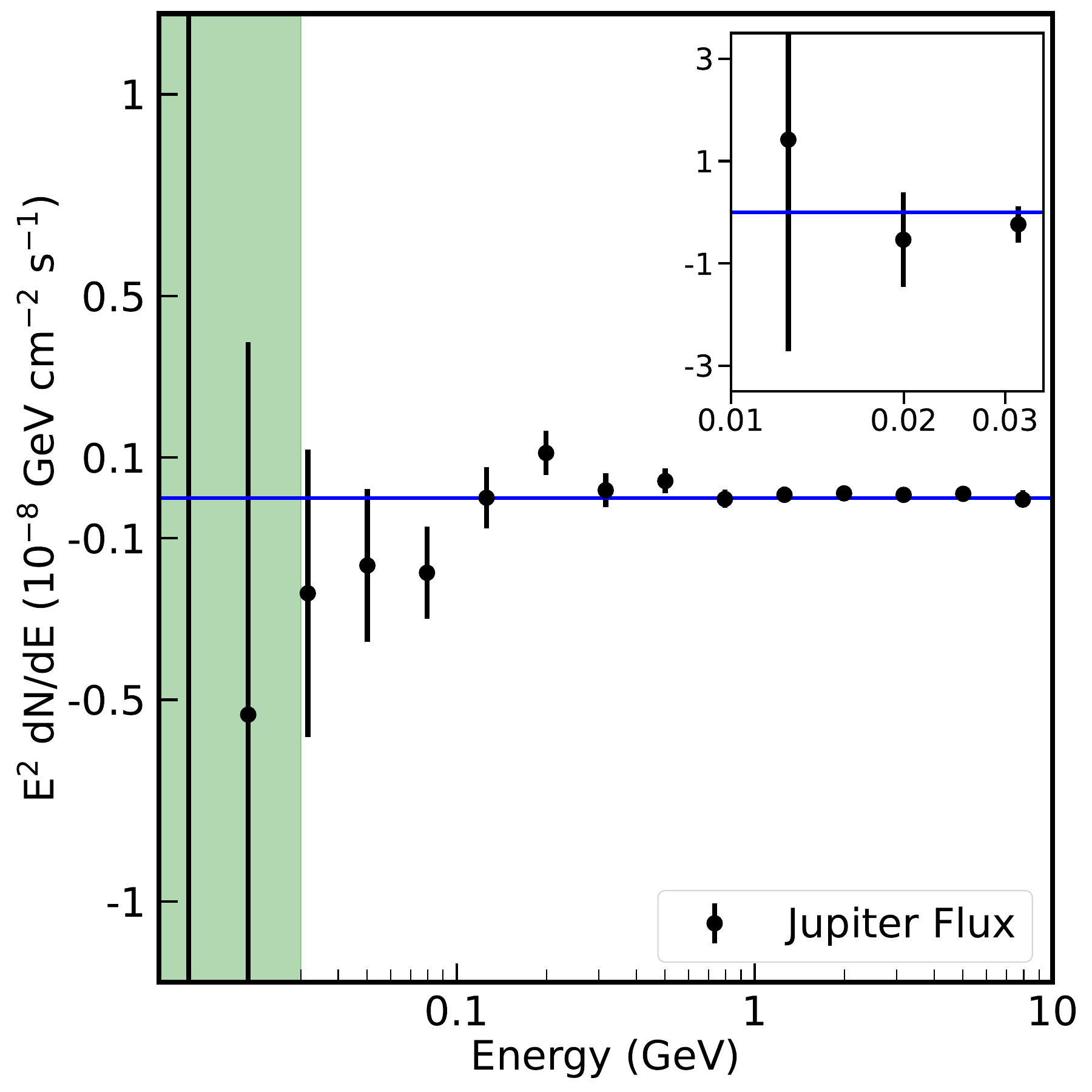}
    \caption{Same as Figure~3 from the main text, except that the results are calculating using either 10 (left) or 5 (right) logarithmic energy bins per decade. The spectral features of the Jupiter data remain the same -- while the amplitude of the positive in negative residuals in individual bins decreases as expected. See text for details.}
    \label{fig:jupiterflux_bins}
\end{figure*}

Figure~\ref{fig:jupiterflux_bins} shows the results of this analysis, noting four important results. First, the spectrum of excesses and deficits in the data is identical for different choices of spatial binning -- implying that our fluctuations in our background model have no effect on the finely binned data. Second, the amplitude of the (statistically insignificant) excesses and deficits decreases in more coarsely binned data. This is expected -- because a spectral energy analysis calculates the fit for a non-physical injected $\gamma$-ray spectrum which has 0 flux below (above) the minimum (maximum) energy of the bin, and a constant flux within the bin. If the energy range of the bin is smaller, then the maximum flux of the bin is larger (for a constant number of total photons in the bin). When the bin is smaller than the energy dispersion (and then the dispersion correctly shifts photons into neighboring bins), the likelihood function will record a higher predicted flux (with an identical likelihood distribution) in the smaller bin. This is the expected behavior of a spectral energy distribution. 

Third, however, the fit and constraints on realistic astrophysical spectra (which span multiple bins) remains unchanged. While stressing that we find no statistically significant detection of Jupiter, we can compare the maximum statistical preference for a Jupiter flux between our analyses. In our default analysis, our model prefers an astrophysical injection spectrum of E$^{-2.1}$ with a total $\gamma$-ray flux between 10~MeV and 10~GeV of 10.0$\times$10$^{-13}$~erg~cm$^{-2}$s$^{-1}$ at the level of 1.45$\sigma$. For our analysis with 10 energy bins per decade, the model prefers an astrophysical injection spectrum of E$^{-2.1}$ with a total $\gamma$-ray flux between 10~MeV and 10~GeV of 9.0$\times$10$^{-13}$~erg~cm$^{-2}$s$^{-1}$ at a level of 1.25$\sigma$. For our analysis with 5 energy bins per decade, the model prefers an astrophysical injection spectrum of E$^{-2.2}$ (by only TS~=~0.02 over E$^{-2.1}$) at a level of 1.31$\sigma$. For the matching E$^{-2.1}$ spectrum, the best-fit flux is 9.4$\times$10$^{-13}$~erg~cm$^{-2}$s$^{-1}$.

Fourth, we note that there is some significant change in the preference for a low-energy excess near 10~MeV. Stressing that we do not believe this excess to be robust, we nevertheless note that such behavior is expected regardless of the existence of the excess. Our finely binned model prefers an extremely soft $\sim$E$^{-3.5}$ spectrum for the excess. However, by combining energy bins, our model assumes a much harder $\sim$E$^{-1}$ spectrum within each bin. Moreover, because the instrumental effective area increases rapidly at low energies (it is 36 times larger at 20~MeV than 10~MeV), the higher-energy data completely dominates any statistical result from the lowest-energy analyses. Thus, while our coarser analyses provides no credence to the low-energy excess, we also do not believe that it invalidates such a result.

\section{Additional Discussion of Astrophysical Implications of our Jovian Gamma-Ray Measurement}
\label{sec:appastro}

In addition to serving as a sensitive probe of dark matter physics, our upper limits on Jovian $\gamma$-rays may have significant implications for particle acceleration in Jupiter's powerful magnetic fields. Jovian magnetic fields are about 20,000 times stronger than Earth's, and power a strong radiation belt that is capable of accelerating protons and electrons to energies as high as 100~MeV~\cite{2017JGRA..122.5148N}. While this would typically produce $\gamma$-ray emission at energies $\lesssim$1~MeV, which falls below the detection limit of current searches, models indicate that the acceleration of particles could be temporally variable~\cite{2017GeoRL..44.4481B}, which may suggest a significant MeV-scale $\gamma$-ray flux during acceleration events.

Perhaps more importantly, galactic cosmic-rays should impinge on Jupiter's surface producing a bright $\gamma$-ray flux. Interestingly, recent observations have found that solar magnetic fields unexpectedly ``turn" a large fraction of incoming cosmic-rays to produce a bright $\gamma$-ray flux at GeV energies that exceeds theoretical predictions by as much as 2~orders of magnitude at high energies~\cite{Linden:2018exo, Tang:2018wqp, Linden:2020lvz, HAWC:2022khj}, approaching an $\mathcal(O)(1)$ efficiency in converting incoming cosmic-rays to outgoing at an energy of $\sim$100~GeV. The physics of this process is unknown. While the small angular size of Jupiter would make such emission fall about a factor of 10 below our current limits at 100 GeV, the low energy $\gamma$-ray emission can be significantly enhanced due to backsplash $\pi^0$'s~\cite{Zhu2023_TBS}, similar to the bright albedo observed from the Earth's surface~\cite{2009PhRvD..80l2004A}. Assuming a similar efficiency to that observed in the Sun, at a lower energy of 1~GeV, would produce a Jovian $\gamma$-ray flux of 10$^{-9}$~GeV~cm$^{-2}$~$s^{-1}$, which is in tension with our observations. Thus, our observations of low-energy Jovian $\gamma$-rays act as a diagnostic of the shielding provided to galactic cosmic-rays by Jovian magnetic fields, which are expected to modulate the cosmic-ray flux at energies below $\sim$14~GV~\cite{2002GeoRL..29.1298S}.

\section{Additional Discussion on Jovian Dark Matter}
\label{sec:appdm}

While capture of DM in celestial objects has been considered in many objects, such as neutron stars and white dwarfs~\cite{Goldman:1989nd,
Gould:1989gw,
Kouvaris:2007ay,
Bertone:2007ae,
deLavallaz:2010wp,
Kouvaris:2010vv,
McDermott:2011jp,
Kouvaris:2011fi,
Guver:2012ba,
Bramante:2013hn,
Bell:2013xk,
Bramante:2013nma,
Bertoni:2013bsa,
Kouvaris:2010jy,
McCullough:2010ai,
Perez-Garcia:2014dra,
Bramante:2015cua,
Graham:2015apa,
Cermeno:2016olb,
Graham:2018efk,
Acevedo:2019gre,
Janish:2019nkk,
Krall:2017xij,
McKeen:2018xwc,
Baryakhtar:2017dbj,
Raj:2017wrv,
Bell:2018pkk,
Garani:2018kkd,
Chen:2018ohx,
Garani:2018kkd,
Dasgupta:2019juq,
Hamaguchi:2019oev,
Camargo:2019wou,
Bell:2019pyc,
Garani:2019fpa,
Acevedo:2019agu,
Joglekar:2019vzy,
Joglekar:2020liw,
Bell:2020jou,
Dasgupta:2020dik,Garani:2020wge,
Leane:2021ihh}, the Sun~\cite{Batell:2009zp,Pospelov:2007mp,Pospelov:2008jd,Rothstein:2009pm,Chen:2009ab,Schuster:2009au,Schuster:2009fc,Bell_2011,Feng:2015hja,Kouvaris:2010,Feng:2016ijc,Allahverdi:2016fvl,Leane:2017vag,Arina:2017sng,Albert:2018jwh, Albert:2018vcq,Nisa:2019mpb,Niblaeus:2019gjk,Cuoco:2019mlb,Serini:2020yhb,Mazziotta:2020foa,Bell:2021pyy}, Earth~\cite{Freese:1985qw,Mack:2007xj,Chauhan:2016joa,Bramante:2019fhi,Feng:2015hja}, Uranus~\cite{Mitra:2004fh,Adler:2008ky}, Neptune and Jupiter~\cite{Adler:2008ky,Kawasaki:1991eu}, Mars~\cite{Bramante:2019fhi}, and even exoplanets~\cite{Leane:2020wob}, detection prospects for DM signatures from Jupiter have not yet been studied with existing data (see also Ref.~\cite{Batell:2009zp} where it was noted detection may be possible with Milagro). We use our new $\gamma$-ray search results to study observable DM signatures in Jupiter for the first time.

In this section, we discuss additional aspects of Jovian DM and this new search with $\gamma$-rays. This includes caveats on the evaporation mass, details of the DM capture and annihilation processes, comparison of our new Jovian search with other searches in this parameter space, spectral assumptions, and lastly model specific interpretations of these results.

\subsection{Dark Matter Evaporation}
\label{sec:evap}

The successful capture of DM by Jupiter is a balance between the kinetic energy transferred to the DM from the Jovian temperature and the DM's potential energy within Jupiter. As the DM becomes lighter, it is easier for it to escape Jupiter, and consequently not produce an annihilation signal. This threshold is called the evaporation mass, and the procedure to calculate it was first outlined in Ref.~\cite{1987ApJ...321..560G}. As emphasized in the main text, the evaporation mass is highly model dependent. As such, we have shown our Jovian gamma-ray sensitivities roughly as low as Fermi thresholds allow, which corresponds to a DM mass $\mathcal{O}$(10 MeV) or higher. For some benchmark model examples, the evaporation mass can be about 200 MeV - 1 GeV in our parameter space for DM with heavy mediators, while it can be as low as at least sub-MeV for DM with light mediators, see Ref.~\cite{Acevedo:2023owd} for detailed discussion. Therefore, we expect that a detailed model treatment of the evaporation mass can produce less sensitive results than the $\mathcal{O}$(10 MeV) Fermi cutoff shown here. As the goal of our work is to demonstrate the strong power of Jovian $\gamma$-rays to probe DM using our new measurements, rather than perform a DM model-dependent study of all possibilities, we leave a detailed study of DM-model specific scenarios with Jupiter to future work.

\subsection{Dark Matter Annihilation}
\label{sec:DM}
 
If DM can self-annihilate, there is an interplay between the capture and annihilation in Jupiter. DM annihilation depletes the incoming captured DM, such that the number of DM particles inside Jupiter $N(t)$ evolves over time, governed by \cite{Kouvaris:2010}
\begin{equation}
\frac{dN(t)}{dt} = C_{\jupiter} - C_A N(t)^2
\label{eq:evolve}
\end{equation}
where $C_{\jupiter}$ is the total capture rate given in Eq.~\ref{eq:multiscatter_total} and $C_A = \langle\sigma_A v\rangle/V$ is the thermally averaged annihilation cross section over the effective annihilation volume. Eq.~\ref{eq:evolve} has the solution
\begin{equation}
    N(t) = \sqrt{\frac{C_{\jupiter}}{C_A} }\tanh \frac{t}{t_{\rm eq}},
\label{eq:nevol}
\end{equation}
where $t_{\rm eq} = 1/\sqrt{C_A C_{\jupiter}}$ is the timescale to obtain an equilibrium between capture and annihilation of DM within Jupiter. Once equilibrium has been reached, the annihilation rate ($\Gamma_{\rm ann}$) is simply
\begin{equation}
    \Gamma_{\rm ann} = \frac{C_{\jupiter}}{2},
\label{eq:equlbm}
\end{equation}
where the factor 1/2 arises because DM annihilation involves two particles. For models with very small capture rates, equilibrium may not be reached within Jupiter's lifetime. However, detectable fluxes are still possible. In this case, the annihilation rate at time $t$ is
\begin{eqnarray}
\Gamma_{\rm ann}(t)=\frac{N(t)^2}{4\,V_{\rm eff}}\,\langle\sigma\,v\rangle,
\label{eq:vol}
\end{eqnarray}
where $\langle\sigma\,v\rangle$ is the thermally averaged DM annihilation cross section, $N(t)$ is the DM number from Eq.~\ref{eq:nevol}, $V_{\rm eff}$ is the effective volume of DM particles in Jupiter. This assumes non-self-conjugate particles; an additional factor of $2$ is required on the right-hand side for self-conjugate particles.

We assume that the mediator escapes the object without attenuation. This assumption is justified, as long-lived particles generally have weak couplings, which can suppress any scattering cross section of the particle (see Ref.~\cite{Leane:2021ihh} for discussion of this point).

\subsection{Comparison with Other Searches}

In models including DM annihilation to long-lived mediators, searches for Jovian $\gamma$-rays provide a superior reach to direct detection searches. In the sub-GeV DM regime, direct detection experiments lose sensitivity as the recoils become increasingly weak, and are eventually below detector thresholds. Jupiter on the other hand, is optimized to search for DM particles with masses of around the proton mass, and has strong sub-GeV DM sensitivity. Note that compared to the Jupiter limits, the direct detection limits do not require any minimum annihilation cross section.

Searches for DM annihilation to long-lived particles in the Galactic Center from a population of brown dwarfs~\cite{Leane:2021ihh} have the most overlap with our new Jovian parameter space. Brown dwarfs in the Galactic bulge are generally old ($\gtrsim\,$Gyr), such that their radii are comparable to that of Jupiter (brown dwarfs cool over time, eventually settling into this common radius). Their capture radius is, however, enhanced by the fact that they are more massive, and benefit from gravitational focusing. A large population of brown dwarfs exist in the Galactic center, where they also benefit from large DM densities compared to the local position. However, despite these many benefits, the Galactic center signal is much weaker due to its large distance, with the expected flux diminishing proportionally to the inverse square of the distance. Jupiter on the other hand, is only one object in a comparably low DM density, but it is very close to Earth. This is the main reason the new Jovian search sensitivities are strong in comparison. Note however that brown dwarfs can reach equilibrium for smaller annihilation cross sections than Jupiter (see equilibrium subsection below), making them more optimal for $p$-wave DM models.

Another nearby, and even bigger object, is the Sun. In Fig.~\ref{fig:limits} we show limits from DM annihilating to long-lived mediators in the Sun, as calculated in Refs.~\cite{Leane:2017vag,Albert:2018jwh}. We see that the solar limits extend to much lower cross sections, owing to the Sun being much larger than Jupiter and closer to the Earth. The Jupiter limits extend to lower DM masses, because Jupiter’s cooler core in part can prevent evaporation of sub-GeV DM. However, depending on the specific DM model of interest, we emphasize that the low-mass end of the sensitivity can substantially change. In the case of only heavy mediators, the DM evaporation mass is around about 200 MeV - 1 GeV for Jupiter depending on the scattering cross section. In the case of light mediators, the DM evaporation mass can be instead at least sub-MeV~\cite{Acevedo:2023owd}.

We note that these indirect detection targets are also optimized for different decay lifetimes/boosts of the mediator. The Solar DM search is not sensitive to as short decay lifetimes/as small boosts as Jupiter, as the Sun is an order of magnitude larger in radius, and the particles must therefore travel further to escape the Sun. The GC brown dwarf search has comparable decay lifetime requirements as Jupiter, but has the advantage of probing a wider range of decay lengths; very \textit{long} decay lengths are still likely to be detectable as there is a large distance between the GC and the \textit{Fermi} Telescope at the local position (see the model subsection for more discussion of model-dependence and interplay of the different constraints in Fig.~\ref{fig:limits}).

A second important difference between the brown dwarf GC limits and Jovian limits, is that the GC limits are highly sensitive to the DM density profile. The brown dwarf limits~\cite{Leane:2021ihh} shown in Fig.~\ref{fig:limits} are for a Navarro-Frenk-White (NFW) density profile~\cite{Navarro:1996}, with an inner slope index of $\gamma=1$. However, it is not currently known if the inner Galaxy exhibits a DM cusp (NFW-like) or core at its center. In this sense, the Jupiter limits are more robust, as the DM density at the solar system position is known to much higher precision. If a cored-DM density profile were the true DM profile, the brown dwarf GC limits would substantially weaken, to be much weaker than our new Jovian constraints. As such, we only show a line for the BD limits; if a NFW profile were observed, cross-sections above this line would be ruled out. If a cored DM density profile were realized, the limits would be much weaker than the line shown.

Finally, we note that we have calculated our Jovian DM limits to 95$\%$ C.L., while the solar and brown dwarf limits take a more conservative approach of requiring the DM flux to exceed the measured flux in any energy bin before any limits are set. The Jovian DM limits extend to lower DM masses than those from Galactic center brown dwarfs. This is simply because the Galactic center $\gamma$-ray dataset had not been analyzed to as low energies as our Jovian $\gamma$-ray analysis. If they had been, the brown dwarf limits would be expected to extend to lower DM masses than those with Jupiter.

\subsection{Equilibrium Assumptions}

We provide some brief estimates on the range of cross sections in Fig.~\ref{fig:limits} which can lead to equilibrium between DM scattering and annihilation in Jupiter. We find that DM in Jupiter reaches equilibrium for thermally averaged annihilation cross sections of $\langle \sigma_{\rm ann} v_{\rm rel}\rangle \gtrsim 10^{-30}\left(m_\chi/\text{GeV}\right)^{-1} \, \rm cm^3/s$, when DM has its geometric scattering cross section in Jupiter of $\sim10^{-32}~$cm$^2$. Therefore, for $s-$wave annihilation cross sections to come into equilibrium (which have $\langle \sigma_{\rm ann} v_{\rm rel}\rangle \sim 10^{-26}\, \rm cm^3/s$), only scattering rates larger than about $\sim10^{-36}~$cm$^2$ will come into equilibrium within the current age of Jupiter (4.5 billion years). However, we emphasize that the equilibration timescale will be much faster if DM thermalizes within Jupiter and settles into an even smaller thermal volume. In this case, lower cross sections can still fall into equilibrium. Thus, in Fig.~\ref{fig:limits} we show the full potential range that can be probed with Jupiter. Note that even if equilibrium is not reached, a detectable $\gamma$-ray signal can still be produced. In general, the exact limits and thermalization process will depend on the particle DM model in question.

\subsection{Other Final States and Upcoming MeV Telescopes}

In Fig.~\ref{fig:limits}, we have only shown direct decay to $\gamma$-rays $\chi\chi\rightarrow\phi\phi\rightarrow4\gamma$ (which is the same as the solar and brown dwarf limits shown). This direct decay to $\gamma$-rays is the most optimistic case, providing the strongest limits. However, other final states can also be probed with our search. For example, with electron, muon, or pion/light quark final states, the DM mass sensitivity can also be extended beyond current solar limits. The sensitivity will be weaker for these final states, because their resulting $\gamma$-ray spectra are softer, such that the $\gamma$-ray flux peaks at much lower energies, where the sensitivity is worse. We nonetheless expect given the ten orders of magnitude potential gain, that even these final states with weakened sensitivity can provide new constraining power. We note that in general, the \textit{Fermi} Telescope's performance is increasingly poor at sub-GeV $\gamma$-ray energies. As such, upcoming MeV telescopes such as AMEGO~\cite{McEnery:2019tcm} and e-ASTROGAM~\cite{DeAngelis:2017gra} may provide more insightful measurements for this lower DM mass range in the near future.

\subsection{Model Interpretation of Results}

\label{sec:models}

We have presented our limits in a model-independent way, requiring only that the mediator is sufficiently boosted or long-lived to escape Jupiter. Our goal is not to perform an extensive study of the model possibilities, but rather to point out the strong potential constraining power for DM annihilation to long-lived particles in Jupiter. Given the vast improvement in parameter space from other experiments, we expect a broad range of models can be accessed with this new Jovian $\gamma$-ray search. We emphasize that, in terms of limits, we show the optimal scenario. Indeed, depending on the particular model realizations and parameters, these limits can be weaker. This applies to both the cross-section sensitivity, and the DM evaporation mass.

We note, in any case, that a number of models are promising for this region of parameter space. In particular, those already pointed out for DM annihilation to long-lived particles in the Sun can produce similar signals, such as a mixed scalar-pseudoscalar mediator model~\cite{Arina:2017sng}, pseudoscalars~\cite{Batell:2009zp}, vectors or dark Higgs bosons~\cite{Batell:2009zp}. For example, for the scalar-pseudoscalar model, we expect combinations of DM-mediator couplings of $\lesssim1$ and proton-mediator couplings $\lesssim10^{-3}$, mediator masses less than about an order of magnitude smaller than the given DM mass, can accommodate our Jovian signal, and be compatible with other constraints (see Ref.~\cite{Arina:2017sng} for details). 

Overall, it is important to keep in mind that the limits shown in Fig.~\ref{fig:limits} have different assumptions, and in a model-dependent context, the interplay of these bounds will vary. For example, while the limits in Fig.~\ref{fig:limits} are focused on the sub-GeV mass regime, it is possible that Jupiter can also be used as a superior target in the GeV and above DM mass range. This can occur, for example, if the decay length is shorter than the Sun radius, such that the solar signal is sufficiently extinguished, leaving weaker or no solar constraints in this higher mass region. 

Furthermore, the direct detection constraints may be weakened for classes of models such as inelastic DM, where there are two dark sector states with a mass splitting. The mass splitting can make the up-scatter process of the lighter to the heavier state highly energetically suppressed for direct detection. However, for Jupiter, as it is a different system with different kinematics, it may be possible to still have capture, and a consequent annihilation signal. It is also possible to have more than one mediator present (and can for example be required to provide dark sector mass generation), which also would change the relative phenomenology~\cite{An:2013yua, Kahlhoefer:2015bea,Bell:2016fqf,Bell:2016uhg,Duerr:2016tmh,Bell:2017irk,Cui:2017juz}. 

Lastly, we note that strongly-interacting DM models naturally reside in the sub-GeV parameter space relevant for Jupiter, particularly for the large cross sections that can be probed. This includes Strongly Interacting Massive Particles (SIMPs)~\cite{Hochberg:2014dra} and Co-SIMPs~\cite{Smirnov:2020zwf}. They also are not constrained by any other observations in the relevant parameter space, including indirect detection and direct detection experiments~\cite{Smirnov:2020zwf}. However, as these classes of models have $3\rightarrow2$ DM annihilation processes, the $\gamma$-ray energy spectrum will be different to the $2\rightarrow2$ annihilation assumption used to calculate limits in this work.

\bibliography{bib}

\end{document}